\documentclass[twocolumn]{aastex631}

\def\ltsima{$\; \buildrel < \over \sim\;$}
\def\ltsim{\lower.5ex\hbox{\ltsima}}
\def\gtsima{$\; \buildrel > \over\sim \;$}
\def\gtsim{\lower.5ex\hbox{\gtsima}}
\def\ms{$M_{\odot}$ }
\def\msp{$M_{\odot}$}

\accepted{to ApJ}

\shorttitle{Supernovae with a Metallicity Threshold}
\shortauthors{Tsujimoto}

\begin{document}

\title{Joint R-process Enrichment by Supernovae with a Metallicity Threshold and Neutron Star Mergers}

\correspondingauthor{Takuji Tsujimoto}
\email{taku.tsujimoto@nao.ac.jp}

\author[0000-0002-9397-3658]{Takuji Tsujimoto}
\affiliation{National Astronomical Observatory of Japan, Mitaka, Tokyo 181-8588, Japan}



\begin{abstract}
The enrichment history of r-process elements has been imprinted on the stellar abundances that change in accordance with increasing metallicity in galaxies. Close examination of the [Eu/Fe] feature caused by stars in nearby galaxies, including the Large Magellanic Cloud (LMC), shows its perplexity. The decreasing trend of the [Eu/Fe] feature is followed by a nearly constant value;~this trend is generally attributed to an onset of the delayed Fe release from type Ia supernovae (SNe Ia), which is the same interpretation of the [$\alpha$/Fe] feature. However, this feature appears in the LMC at [Fe/H] of approximately -0.7, which is significantly higher than that for the [$\alpha$/Fe] case ($\approx$ -2). This result potentially indicates the presence of an overlooked property of the r-process site that remains unseen in the study of the Milky Way. Here, we propose that this [Eu/Fe]-knee feature is created by a fade-out of core-collapse SNe producing r-process elements;~these elements along with neutron star mergers (NSMs) promote the r-process enrichment under the condition for this specific SNe such that their occurrence is limited to a low-metallicity environment. This metallicity threshold for the occurrence rate of r-process SNe at a subsolar is nearly identical to that for long gamma-ray bursts whose origin may be connected to fast-rotating massive stars. Moreover, we reason that the contribution of Eu from NSMs is crucial to maintain a high [Eu/Fe] at an early stage in dwarf galaxies by a balance with Fe from SNe Ia;~both enrichments via NSMs and SNe Ia proceed with similar delay time distributions.
\end{abstract}


\section{Introduction}

Currently, at least two major updates to our understanding of the r-process have occurred through ongoing research after the discovery of gravitational waves from the neutron star merger (NSM) GW170817. First, spectral analyses of its kilonova, i.e., an electromagnetic counterpart (AT2017gfo) are used to identify the features corresponding to some specific r-process elements \citep[e.g.,][]{Watson_19, Domoto_22}. These findings leave minimal doubt regarding whether NSMs are r-process sites \citep[see also the latest JWST result;][]{Levan_24}. Second, Galactic chemical evolution does not appear to be reconciled with the sole site of the r-process by NSMs since their relatively long delay time for r-process enrichment cannot  explain the stellar record of the abundances of these products in the Galactic stars \citep[e.g.,][]{Cote_19, Siegel_19, Tsujimoto_21}. These deductions lead the presence of another production site whose delay time is quite short and can be identified as some stellar collapses of massive stars, such as magnetorotational supernovae (SNe) \citep[e.g.,][]{Winteler_12, Nishimura_15} or collapsars \citep[e.g.,][]{Siegel_19, Li_23} (hereafter, these are referred to as r-process SNe).

The best stellar record tracing the time evolution of the r-process element is the correlation of [Eu/Fe] with [Fe/H] for individual stars in the galaxies. This evident feature can be approximated by a plateau-like [Eu/Fe] ratio at low metallicities and the subsequent decreasing trend with increasing [Fe/H], which causes a transition point in the [Eu/Fe]-knee feature. This overall feature is identical to that for [$\alpha$/Fe] \citep[e.g.,][]{Sneden_08, Tolstoy_09}, and the decreasing trend of these abundance ratios is theoretically caused by delayed Fe release from type Ia supernovae (SNe Ia). 

\begin{figure}[t]
	\vspace{0.3cm}
    \hspace{0.8cm}
	\includegraphics[width=0.77\columnwidth]{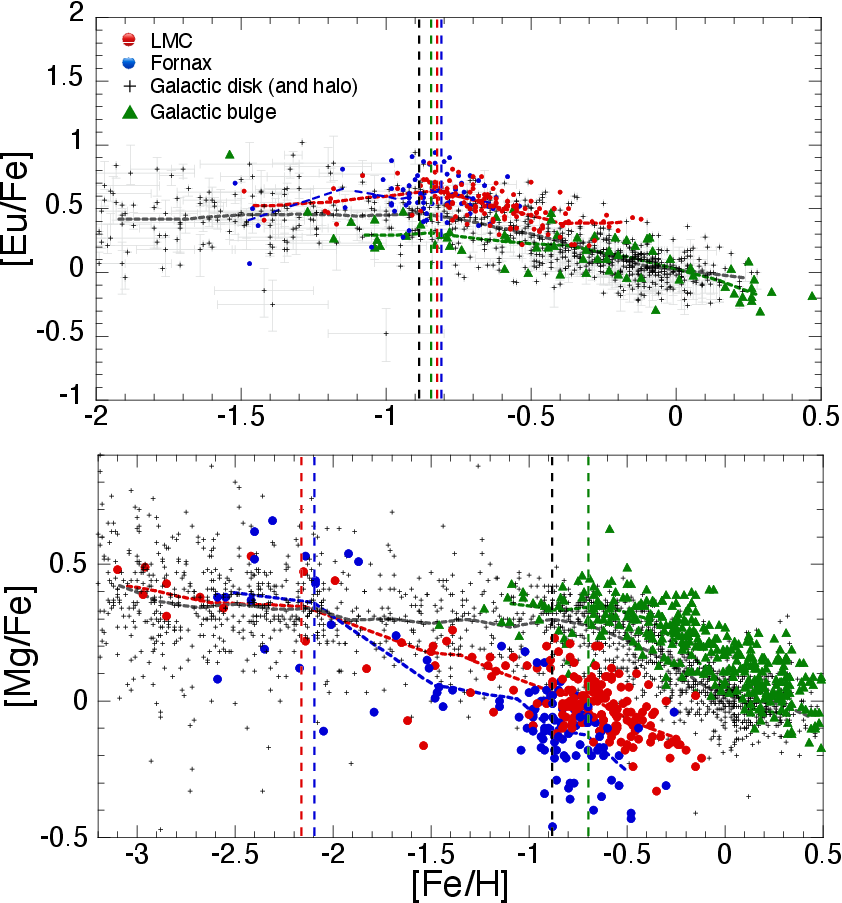}

\caption{Top panel:~Observed correlation of [Eu/Fe] with [Fe/H] for the LMC \citep[red circles:][]{Swaelmen_13}, the Fornax dSph galaxy \citep[blue circles:][]{Letarte_10}, solar neighborhood stars (i.e., Galactic disk and halo stars;~pluses: the data from the SAGA database), and Galactic bulge stars \citep[green triangles:~the data assembled by][]{Tsujimoto_19}. The positions of the [Eu/Fe] knee for each dataset are estimated by the mean evolutionary change in [Eu/Fe] with increasing [Fe/H]. Both the knee positions and the mean curves are indicated by the dashed lines with the same color of the data. Bottom panel:~Same as top panel but for [Mg/Fe]. Additional observed data are plotted for the LMC \citep{Reggiani_21, Chiti_24} and the Fornax dSph galaxy \citep{Letarte_06, Lemasle_14}.
}
\end{figure}

However, thus far, the accumulated abundance data beyond the solar neighborhood challenge the theoretical deduction on how the [Eu/Fe] knee appears. The arising problem is demonstrated in Figure 1. The top panel of this figure shows the correlation of [Eu/Fe] with [Fe/H] for the Galactic components of the disk and bulge and for the relatively massive nearby dwarf galaxies of the Large Magellanic Cloud (LMC) and the Fornax dwarf spheroidal (dSph) galaxy\footnote{Less massive dwarf galaxies, such as the Draco dSph galaxy, show a different Eu/Fe feature, which is characteristic of a decreasing [Eu/Fe] (a constant [Eu/H]) and not dependent on the metallicity;~this is an outcome of no occurrence of an r-process event associated with its rarity and a continuous release of Fe from canonical core-collapse SNe (CCSNe)  \citep[e.g.,][]{Tsujimoto_14, Ji_16a, Reichert_20, Molero_21}. The same reasoning also applies to [Ba/Fe] ([Ba/H]) in ultra-faint dwarf galaxies \citep[e.g.,][]{Frebel_14, Tsujimoto_14, Ji_16b}.}. Intriguingly, minimal difference is observed in the positions of the [Eu/Fe] knees among the four distinct populations, which are formed with a largely different speed of star formation/chemical enrichment from the fastest (the bulge) to the slowest (the Fornax dSph galaxy). This result strongly indicates that the [Eu/Fe] knee was not caused by an onset of the Fe release from SNe Ia, since the metallicity for the initiation of SN Ia enrichment changes in accordance with the age-[Fe/H] relationship that differs among the Galactic components and galaxies. However, as generally expected, the correlation of [Mg/Fe] with [Fe/H] shows a clear sequence of different [Mg/Fe]-knee positions in order of the enrichment speed (the bottom panel of Fig.~1).

In the following section, based on this reasoning, we assess the production sites of the r-process elements together with their associated properties, which lead to a complete understanding of the elemental features. Then, by incorporating the acquired knowledge into the models, we perform calculations for the chemical evolution for the LMC and the Galactic disk (Section 3) and summarize our conclusions in Section 4.
 
\section{Insights into the sites of r-process production}

Since the [Eu/Fe]-knee locus solely depends on the metallicity, and the metallicity corresponds to [Fe/H]$\approx$-0.7, this feature is  likely associated with the properties of the r-process production site. As the most plausible case, the occurrence rate of the astrophysical phenomena related to r-process production can have a metallicity threshold above which its rate is significantly reduced. This metallicity threshold ($Z_{\rm th}$) at a subsolar can be reminiscent of long gamma-ray bursts (LGRBs) whose formation is biased toward a low-metallicity environment \citep{Fruchter_06}, such as $Z_{\rm th}$=0.3-0.5 $Z_\odot$ \citep{Vergani_15}. If we assume a higher $\alpha$/Fe ratio relative to a solar value such as [$\alpha$/Fe]$\approx0.3$ for LGRB host galaxies, the obtained $Z_{\rm th}$ is approximately equivalent to [Fe/H]$_{\rm th}\approx$ -0.7, which is deduced from the relationship:~[Z/H] = [Fe/H] + $A$[$\alpha$/Fe] with  $A\approx1$ \citep{Thomas_03}. A similar result is also obtained for S abundance in LGRB afterglows:~[S/H]$_{\rm th}\approx$-0.6 \citep[see][]{Wet_23}.

\begin{figure}[t]
	\vspace{0.3cm}
    \hspace{0.5cm}
	\includegraphics[width=0.85\columnwidth]{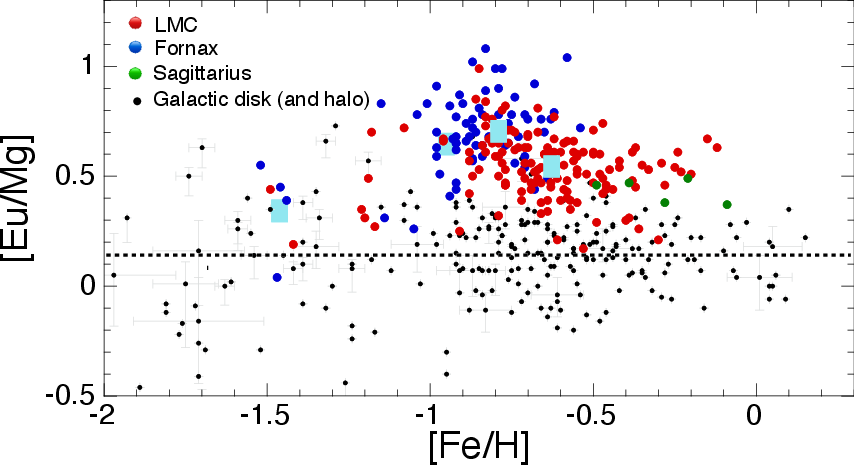}

\caption{Observed correlation of [Eu/Mg] with [Fe/H] for the LMC (red circles), the Fornax dSph galaxy (blue circles), and solar neighborhood stars (black circles). The data are taken from the same sources as in Figure 1. The observed data for the Sagittarius dSph galaxy are also attached \citep[green circles:][]{McWilliam_13}. The light blue symbols indicate a mean [Eu/Mg] value for the LMC and the Fornax dSph galaxy for several metallicity bins with a 0.2 dex width.
}
\end{figure}

The progenitors of LGRBs are proposed to be fast-rotating massive stars ending with collapsars \citep{Woosley_06, Yoon_06} that are  capable of providing the physical conditions necessary to achieve the r-process \citep{Siegel_19}. From a theoretical perspective, a metallicity threshold for LGRBs should exist that retains enough angular momentum as $Z_{\rm th}<$0.3$Z_\odot$, which is reconciled with [Fe/H]$_{\rm th}\approx$ -0.7. Therefore, it is natural to presume that collapsars are associated with the r-process enrichment for [Fe/H]\ltsim -0.7. In addition, magnetorotational SNe may have a similar metallicity threshold and thus be another candidate in parallel to 
collapsars\footnote{From the aspect of interpretation for the Eu/Fe ratio at low metallicities in dwarf galaxies, magnetorotational SNe can be favored over collapsars, as discussed in the next section.}, since they also originate from fast-rotating massive stars and may have a potential connection to superluminous SNe, which are inclined to emerge in low-metallicity galaxies \citep{Lunnan_14, Perley_16, Schulze_18}. Accordingly, we regard the [Eu/Fe]-knee feature, commonly observed in Galactic components and dwarf galaxies, as evidence for the Eu enrichment by r-process SNe \citep[e.g.,][]{Cote_19, Siegel_19, Haynes_19, Reichert_20, Molero_21, Tsujimoto_21}.

Moreover, we also identify a signature of the r-process enrichment by NSMs in the elemental features of dwarf galaxies. This is demonstrated in Figure 2, which shows the correlation of [Eu/Mg] with [Fe/H]. We observe a gradual increasing trend of [Eu/Mg] with increasing [Fe/H] for [Fe/H]\ltsim -0.8, which cannot be explained by a single Eu site from the r-process SNe but needs an additional contribution of a delayed Eu release from NSMs compared to the fast ejection of Mg by CCSNe \citep[see also][]{Tarumi_21}. Since the SN Ia enrichment begins at [Fe/H] \ltsim -1.5, as implied from the [Mg/Fe] feature, a natural interpretation for an almost constant [Eu/Fe] value at low metallicities in the LMC (Fig.1) can be reasoned from the deduction that the Eu enrichment by NSMs is balanced with the Fe enrichment by SNe Ia, both of which proceed with similar delay time distributions (DTDs) approximated by the form of $\propto t^{-1}$ \citep[e.g.,][]{Paterson_20, Maoz_14}.  

In summary, the correlations of [Eu/Fe] and [Eu/Mg] with respect to [Fe/H] for dwarf galaxies indicate that r-process enrichment by NSMs and r-process SNe proceed in parallel at an early galactic evolution, and then its joint enrichment comes to an end owing to a metallicity threshold for r-process SNe at [Fe/H]$\approx$-0.7, which causes the well-known [Eu/Fe]-knee feature.

\section{Chemical evolution of the LMC and the Galactic disk}

Based on the deductions presented in Section 2, we construct models of the chemical evolution for the LMC together with for the Galactic disk. We highlight the reproduction of the observed elemental feature for the LMC;~however, its simultaneous reproduction with the Galaxy case is extremely important.

\subsection{Modeling of r-process enrichment}

We consider two production sites of the r-process, i.e., the NSMs and r-process SNe. We assume that NSMs occur at a rate of one per 300 CCSNe and that the ejected mass of Eu from each event is $2\times10^{-5}$\ms \citep{Tsujimoto_21}. The mass range of the NS progenitors is assumed to be 10-25 \ms by adopting the boundary mass between NS and black hole (BH) for the progenitor star's end at 25 \ms \citep{Janka_12}. For the DTD of NSMs, we use DTD $\propto t_{\rm delay}^{-0.9}$, with a range of $0.03\leq t_{\rm delay}\leq10$ Gyr. Here, we assign its power index based on the DTD for short GRBs of approximately -1 \citep{Paterson_20} and adjust it to a slightly smaller value than that for SNe Ia (refer to the discussion in Section 3.3). For the production rate and yield of r-process SNe, we use the results of magnetorotational SNe deduced from the chemical evolution \citep{Tsujimoto_15} and nucleosynthesis calculations \citep{Nishimura_15};~the Eu ejection is assumed to occur from massive stars forming NSs, i.e., 10-25 \ms stars, at a rate of one per 150 CCSNe, with an ejected Eu mass of $1.2\times10^{-5}$\ms from each SN. Note that if the r-process SNe are collapsars, a smaller rate and a higher yield can be assigned \citep[cf.][]{Siegel_19}. Then, as a key model ingredient, we introduce the metallicity threshold at [Fe/H]=-0.7, beyond which no r-process SNe is assumed to occur.

\subsection{Galaxy models}

We calculate the gas fraction and the abundance of heavy-elements (Mg, Fe, and Eu) in the gas for each galaxy, which was formed through a continuous low-metal infall of material from outside (i.e., the intergalactic medium). The star formation rate is assumed to be proportional to the gas fraction with a constant coefficient of $1/\tau_{\rm SF}$, where $\tau_{\rm SF}$ is a timescale for star formation. For the infall rate, we use the formula that is proportional to $\exp(-t/\tau_{\rm in})$ with an infall timescale of $\tau_{\rm in}$. For the DTD of SNe Ia, we assume DTD $\propto t_{\rm delay}^{-1.1}$, with a range of $0.1\leq t_{\rm delay}\leq10$ Gyr \citep{Totani_08, Freundlich_21}. The nucleosynthesis yields for CCSNe and SNe Ia are taken from \citet{Tsujimoto_95} and \citet{Iwamoto_99}, respectively. Regarding the ones for CCSNe, we slightly modify the theoretical yields to reproduce the observed plateau of abundance ratio (i.e, [Mg/Fe]) for Galactic halo stars \citep[see][]{Shigeyama_98}.

{\it The Galactic disk.} This is composed of two populations:~thick and thin disks. In this study, we solely focus on modeling the thin disk to which a large fraction of disk stars belong. We assume a moderate star formation lasting for 8 Gyrs, given by ($\tau_{\rm SF}$, $\tau_{\rm in}$) = (2.5, 5) in units of Gyr. For the estimate of the amounts of Fe and Mg from canonical SNe, we assume that CCSNe emerge  from 10-50 \ms stars obeying the Salpeter initial mass function (IMF) and that 8 percent (=$f_{\rm Ia}$) of the primary stars in binaries with initial masses in the range of 3-8 \ms explode as SNe Ia. As the initial abundances, we assume a high value of [Fe/H] (=0.2) along with [Mg/Fe]=0 as the thick disk's remaining gas, since the plausible abundances of metal-richest thick disk stars could be [Mg/H]=[Fe/H] $\approx$ 0.2 \citep[see][]{Tsujimoto_19}. Here, we regard the thick disk as the first disk, which was heated up by an ancient merger, such as Gaia-Enceladus \citep{Helmi_18};~this was subsequently followed by the gradual formation of a secondary disk, i.e., the thin disk \citep{Tsujimoto_19, Spitoni_19}.

{\it The LMC.} The star formation is assumed to slowly proceed, parameterized by ($\tau_{\rm SF}$, $\tau_{\rm in}$) = (5, 10) for the duration of 12 Gyrs. This hypothesis may not be a good approximation for the last 2 Gyr if we consider a reported feature of several bursts of star formation for this period \citep{Harris_09}. However, chemical evolution proceeds up to [Fe/H] of approximately -0.5 around 2 Gyr ago \citep{Povick_23}. Therefore, the enrichment trajectories beyond the knee positions for both [$\alpha$/Fe] and [Eu/Fe] are hardly influenced by the potential bursting mode of star formation. Note, on the other hand, that this ingredient gives a large impact on the chemical feature of the LMC at high [Fe/H] including the present \citep{Bekki_12, Vasini_23}.

In addition, we consider another modification of the model input from the Galaxy:~the reduced amounts of Mg and Fe from SNe. First, to be reconciled with the observed higher Eu/Fe ratio of metal-poor LMC stars compared to those for Galactic stars, we reduce the contribution of the two elements from CCSNe by using an IMF lacking very massive stars of $>$25 \msp. The suppression of the formation of massive stars in low star formation galaxies is indicated from both observational and theoretical studies \citep[e.g.,][]{Pflamm_09, Meurer_09, Lee_09}. In the actual conditions, it is possible to consider that this assumed reduction in heavy elements for chemical enrichment could be, at least in part, due to their loss by galactic winds which are likely to occur in dwarf galaxies \citep[e.g.,][]{Matteucci_85, Bradamante_98}. Then, to increase the Eu/Fe (Mg) ratio, we assume that this truncated IMF barely changes the total Eu yield from the r-process SNe. This hypothesis may be acceptable as a possible case if we consider an unidentified mass range of progenitor stars for the r-process SNe. However, this aspect favores magnetorotaonal SNe, which leave NSs rather than BHs, forming collapsars\footnote{Massive stars usually form either NS or BH, and in general, more massive stars are inclined to form BHs. For instance, if the boundary between NS and BH for the progenitor stars' fate exist at approximately 25 \ms \citep[e.g.,][]{Janka_12}, the truncated IMF applied to the LMC hardly influences the total amount of the r-process production. Additionally, there are models for magnetorotational SNe whose progenitor masses are 20-25 \ms \citep[e.g.,][]{Mosta_14, Kuroda_20}. However, there is a claim that collapsars can be formed from stars as low as 10 \ms \citep{Aguilera_20}.}. 

Second, we use a smaller $f_{\rm Ia}$ of 4 percent, i.e., half of that for the Galactic disk, which is indicated from the observed elemental feature of the LMC, particularly an unusually high Ba/Fe ratio under the effective enrichment by SNe Ia \citep{Bekki_12}. Indeed, a different IMF from the Galactic disk's could be another option to generate a low SN Ia rate. However, a smaller $f_{\rm Ia}$ should be a probable solution, if we consider a demand for reduction in relative contribution of Fe from SNe Ia, compared to both Ba from low-mass \citep[\ltsim 3\msp;][]{Gallino_98} AGB stars and Eu from massive (\gtsim 10 \msp) stars, to chemical enrichment in the LMC. 

\begin{figure}[t]
	\vspace{0.3cm}
    \hspace{0.7cm}
	\includegraphics[width=0.77\columnwidth]{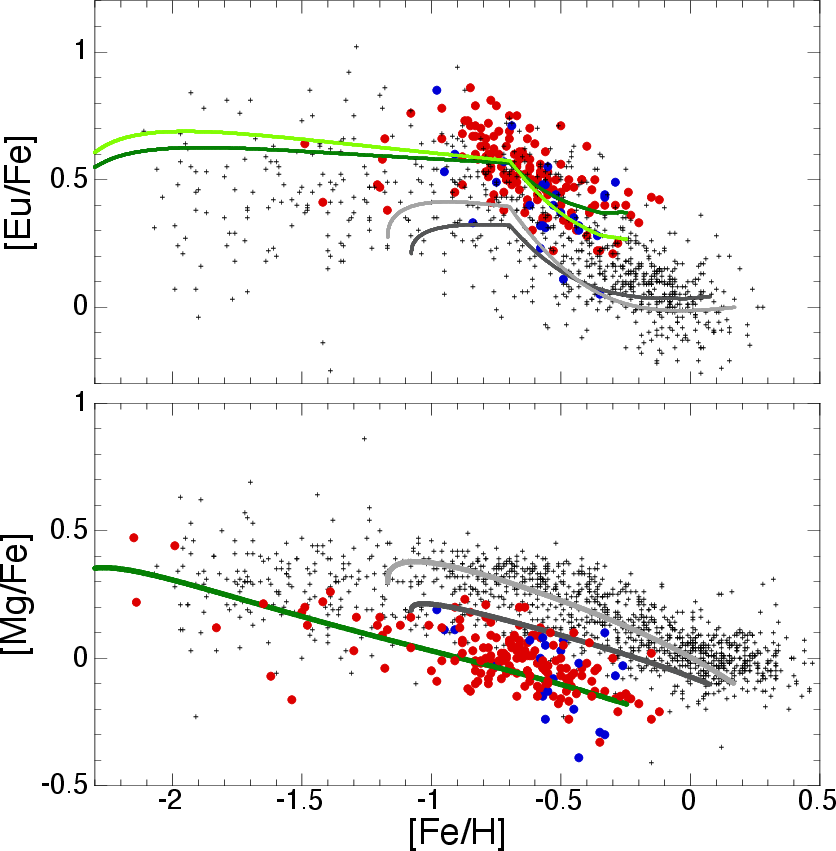}

\caption{Top panel:~Calculated correlations of [Eu/Fe] with [Fe/H] for the LMC (green curves) and the Galactic disk (gray curves) compared to the observed data. For the LMC, the observed data by \citet{Minelli_21} are newly added with blue circles. The dark curves of each galaxy are those deduced from the fiducial models, and the light curves are the modified curves (see the text). Bottom panel:~Same as top panel but for [Mg/Fe]. 
}
\end{figure}

\subsection{Results}

First, from Figure 3, the LMC models successfully reproduce the observed features of both [Eu/Fe] and [Mg/Fe] (green curves). Here, in our modeling, the decreasing trend of [Mg/Fe] with an early start from [Fe/H]$<$-2 is caused by the SN Ia enrichment for Fe, and the Eu/Fe evolution shows its decreasing trend to a metallicity threshold for the r-process SNe at [Fe/H]$\approx$-0.7. It is possible to claim that a steeper decreasing trend of Eu/Fe is more favored by the observation, and the corresponding result can be obtained by changing the relative contribution to Eu between the NSMs and r-process SNe in the models. Then, we calculate one case that consideres an increase in the relative contribution of the r-process SNe by 25\% compared to the fiducial case, and its result is shown by the light green curve in Figure 3.

The model results for the Galactic disk are shown by gray curves in Figure 3. At very early times, [Fe/H] decreases with time owing to dilution by metal-poor infalling gas. Since this reverse evolution ends quickly and thus leaves few stars on this track, the corresponding result from the figure is excluded. In addition, we consider the improved scheme of Galactic dynamics, which suggests that stars radially move on the disk when they encounter transient spiral arms that are naturally generated during the process of disk formation \citep[e.g.,][]{Sellwood_02, Roskar_08, Schonrich_09}. This so-called radial migration of stars predicts that the stars in the solar vicinity represent the mixture of stars born at various Galactocentric distances over the disk. In particular, this dynamic process induces a major migration from the inner disk, which forms faster and becomes more metal-rich than the solar vicinity according to the inside-out scenario \citep{Chiappini_01}. Thus, it is plausible to assume that an in situ chemical enrichment contribute to only a part of the local Galactic chemistry and that the remaining composition must be due to more efficient enrichment trajectories. Here, we calculate the inner-disk \citep[its Galactocentric distance of $\approx$ 4 kpc:][]{Tsujimoto_23} trajectory, by adopting short time scales:~($\tau_{\rm SF}$, $\tau_{\rm in}$) = (0.5, 1). This resultant path (the light gray curve) along with the fiducial path effectively matches the observed data.

Then, the relative behavior of two Eu/Fe paths for [Fe/H]\gtsim -0.5 is more closely analyzed:~a lower [Eu/Fe] value for the fiducial model is reversed and becomes higher with increasing [Fe/H] as an outcome of the Eu enrichment with long delay times by NSMs. This reverse phenomenon is indicated by the abundance analysis of solar twins, and the reported power index of the DTD for NSMs assessed from Galactic chemical evolution is $n$\gtsim 0 \citep{Tsujimoto_21}. However, without contributions from the r-process SNe in a relatively metal-rich regime, as deduced by our study, the required index value is significantly reduced to $-0.9$, i.e., slightly flatter than that (-1.1) for SNe Ia. Our renewed result is reconciled with the measured implications \citep{Zheng_07, Fong_13, Paterson_20}.
 
\begin{figure}[t]
	\vspace{0.3cm}
    \hspace{0.5cm}
	\includegraphics[width=0.85\columnwidth]{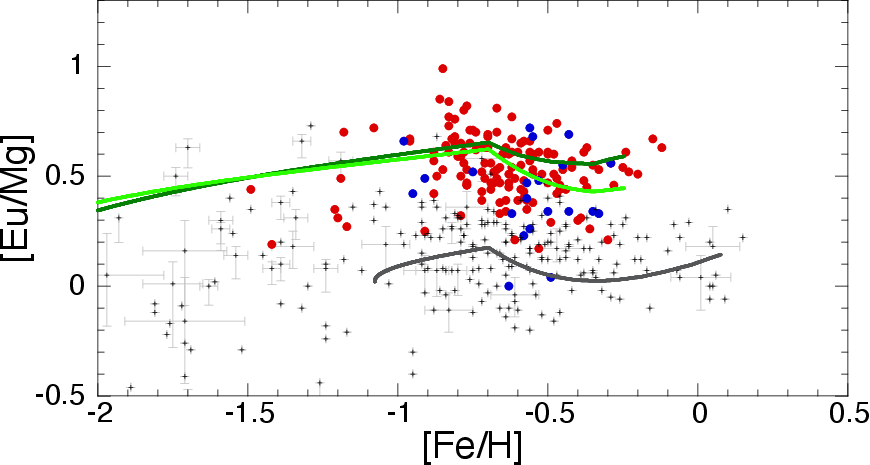}

\caption{Calculated correlations of [Eu/Mg] with respect to [Fe/H] for the LMC (green curves) and the Galactic disk (gray curve) compared to the observed data. Symbols are the same as those in Figure 3.
}
\end{figure}

The contribution of Eu enrichment from NSMs is more visible as an increasing [Eu/Mg] trend for [Fe/H]\ltsim -0.7 for the LMC model (Fig.~4). Afterward, its increasing trend ends and begins to decrease. This wave-like feature is also predicted to occur for the Galactic disk and should be common to a certain extent among galaxies whose star formation has lasted for Gyrs, although in some cases, it is almost invisible owing to a scatter in the observed data.

Finally, we compare the model predictions for the frequency of SN and NSM with the observations. This comparison could lead us to the evaluation of how robust our modeling is. For the Galaxy, the calculated current rates of CCSNe, SNe Ia, and NSMs are 2.23 century$^{-1}$, 0.64 century$^{-1}$, and 33.1 Myr$^{-1}$, respectively. These are in good agreement with the measured rates:~2.30$\pm$0.48, 0.54$\pm$0.12 \citep{Li_11}, and 27.5$^{+42.3}_{-20.6}$ \citep{Abbott_21}. On the other hand, our obtained CCSN+SN Ia rate of 0.14 century$^{-1}$ for the LMC is smaller than the observed value of 0.86$\pm$0.25 century$^{-1}$ \citep{Maoz_10}. That may suggest the presence of bursting star formation in recent years, while the result is broadly consistent with the predicted values (0.064-0.28 century$^{-1}$) by \citet{Vasini_23}.

\section{Summary}

We present two observational features of the [Eu/Fe] knee;~specifically, a switch is observed from its constant value to a decreasing trend with increasing [Fe/H]. This appears (i) at much lower [Fe/H] than their position for their [Mg/Fe] knee in the LMC and the Fornax dSph galaxy and (ii) at similar [Fe/H] values among the dwarf galaxies and the Galaxy, including its bulge component. According to these results, we propose that the [Eu/Fe] knee can be evidence for a metallicity threshold beyond which the SNe producing the r-process elements, referred to as the r-process SNe in this study, cease to emerge. This threshold at [Fe/H]$\approx$-0.7 can be identified for long GRBs, which share their origin of fast-rotating massive stars with the r-process SNe.

In our proposed scenario, the r-process enrichment by NSMs follows the exponential form of DTD with a power index of approximately -1,  similar to that for SNe Ia, and is compatible with the observed [Eu/Fe] feature, as opposed to some chemical evolution studies thus far \citep[e.g.,][]{Cote_19, Tsujimoto_21}. This renewed outcome stems from our hypothesis that the occurrence of the r-process SNe is limited to low-metallicity conditions. 

To more effectively reproduce the observed feature in the LMC, less efficient enrichment by both CCSNe and SNe Ia compared to that for the Galaxy is needed. These reductions in heavy elements may be a result of the IMF lacking very massive (\gtsim 25 \msp) stars and a low frequency of SNe Ia in the dwarf galaxies. This less efficient chemical enrichment is potentially supported by the observed stellar mass-metallicity relationship for galaxies that show a steeper slope for less massive galaxies with an LMC mass scale than that for normal galaxies, including our own \citep{Tremonti_04}.

\begin{acknowledgements}
The author gratefully acknowledges the anonymous referee for providing the comments that improved the work. This work was supported by JSPS KAKENHI Grant Numbers 18H01258, 19H05811, and 23H00132.
\end{acknowledgements}

\end{document}